\documentclass[conference,a4paper]{APSIPA2021}
\usepackage{amsmath, amssymb}
\usepackage{graphicx}
\usepackage{multirow}
\usepackage{threeparttable}
\usepackage{cite}

\usepackage{geometry}
\begin{document}

\renewcommand{\thefootnote}{\fnsymbol{footnote}}

\title{EAViT: External Attention Vision Transformer for Audio Classification}

    
  

\author{
\authorblockN{
Aquib Iqbal\authorrefmark{2}\authorrefmark{1}, 
Abid Hasan Zim\authorrefmark{2}\authorrefmark{1},  
Md Asaduzzaman Tonmoy\authorrefmark{4},
Limengnan Zhou\authorrefmark{5},\\
Asad Malik\authorrefmark{6},
Minoru Kuribayashi\authorrefmark{8}
}

\authorblockA{
\authorrefmark{2}
Department of Computer Science, University of Massachusetts, Amherst, MA, USA,
E-mail: aquibiqbal@umass.edu
}
\authorblockA{
\authorrefmark{3}
Department of Mechanical Engineering, Aligarh Muslim University, India,
E-mail: abid@zhcet.ac.in
}
\authorblockA{
\authorrefmark{4}
Department of Electrical Engineering, Aligarh Muslim University, India,
E-mail: asaduzzaman.tonmoy47@gmail.com
}
\authorblockA{
\authorrefmark{5}
School of Electronic and Information Engineering, UESTC, China,
E-mail: dreamzlmn@foxmail.com
}
\authorblockA{
\authorrefmark{6}
School of Information Technology, Monash University Malaysia, Malaysia,
E-mail: asad.malik@monash.edu
}
\authorblockA{
\authorrefmark{8}
Center for Data-driven Science and Artificial Intelligence, Tohoku University, Japan, 
E-mail: kminoru@tohoku.ac.jp}}

\maketitle

\begin{abstract}

This paper presents the External Attention Vision Transformer (EAViT) model, a novel approach designed to enhance audio classification accuracy. As digital audio resources proliferate, the demand for precise and efficient audio classification systems has intensified, driven by the need for improved recommendation systems, and user personalization in various applications, including music streaming platforms and environmental sound recognition. Accurate audio classification is crucial for organizing vast audio libraries into coherent categories, enabling users to find and interact with their preferred audio content more effectively. In this study, we utilize the GTZAN dataset, which comprises 1,000 music excerpts spanning ten diverse genres. Each 30-second audio clip is segmented into 3-second excerpts to enhance dataset robustness and mitigate overfitting risks, allowing for more granular feature analysis. The EAViT model integrates multi-head external attention (MEA) mechanisms into the Vision Transformer (ViT) framework, effectively capturing long-range dependencies and potential correlations between samples. This external attention (EA) mechanism employs learnable memory units that enhance the network's capacity to process complex audio features efficiently. The study demonstrates that EAViT achieves a remarkable overall accuracy of 93.99\%, surpassing state-of-the-art models.
\footnotetext[1]{Equal contributions.}


\end{abstract}


\section{Introduction}

The increasing availability of digital music resources has led to rapid advancements in multimedia technology. As a result, consumers have progressively shifted towards accessing music through online streaming platforms. AI methods are particularly effective in addressing the complex task of meeting diverse and intricate music retrieval needs from extensive music collections. For instance, consumers may desire to listen to a specific song within a particular genre that conveys a certain emotional tone. In these instances, accurate music labelling is crucial to ensure the precise music is delivered~\cite{elbir2020music, li2023few}. 
\begin{figure}[h!]
  \centering
  \includegraphics[width= 0.45\textwidth]{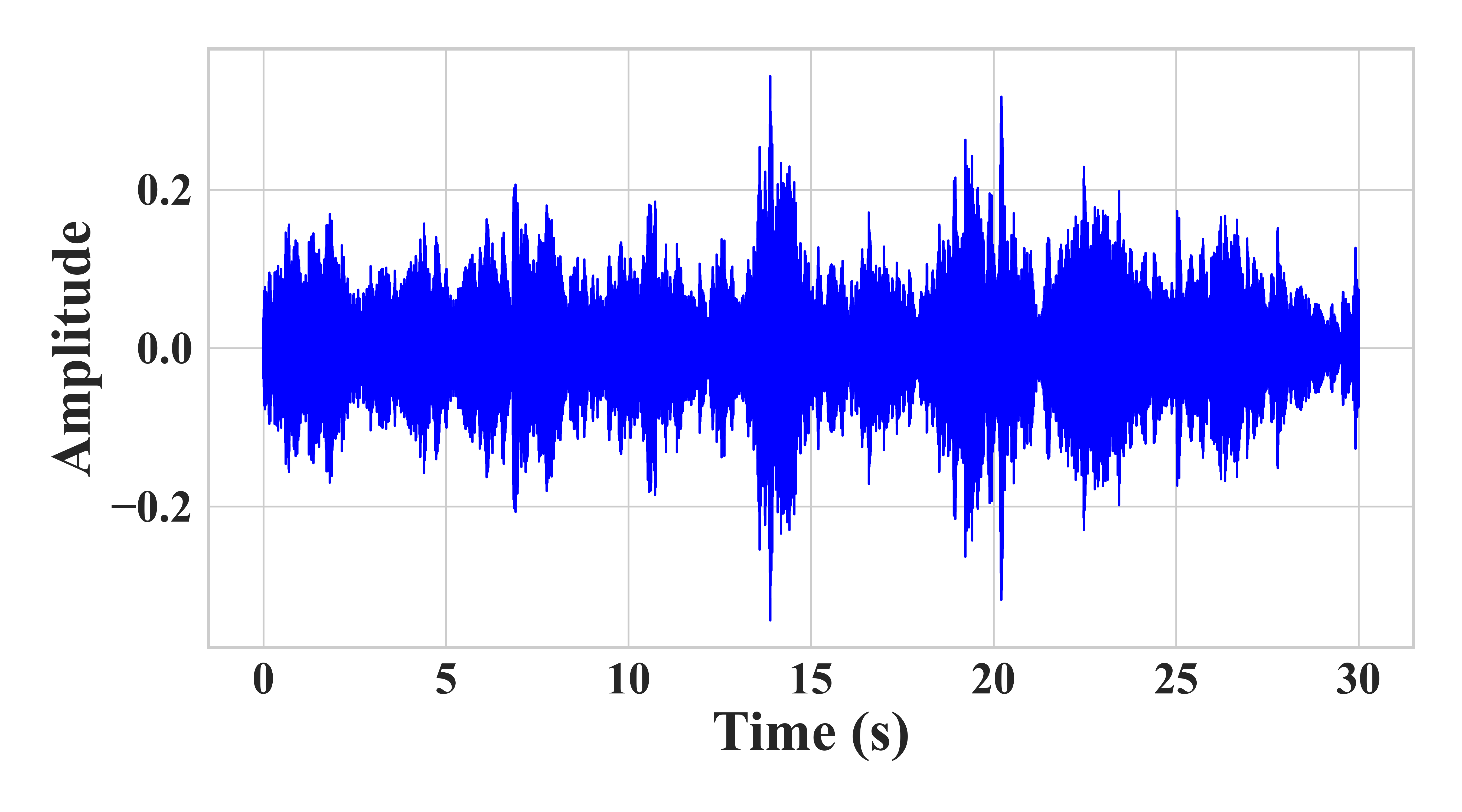}
  \caption{Raw waveform representation of an audio sample.}
  \label{fig:raw}
\end{figure}
\begin{figure}[h!]
  \centering
  \includegraphics[width= 0.46\textwidth]{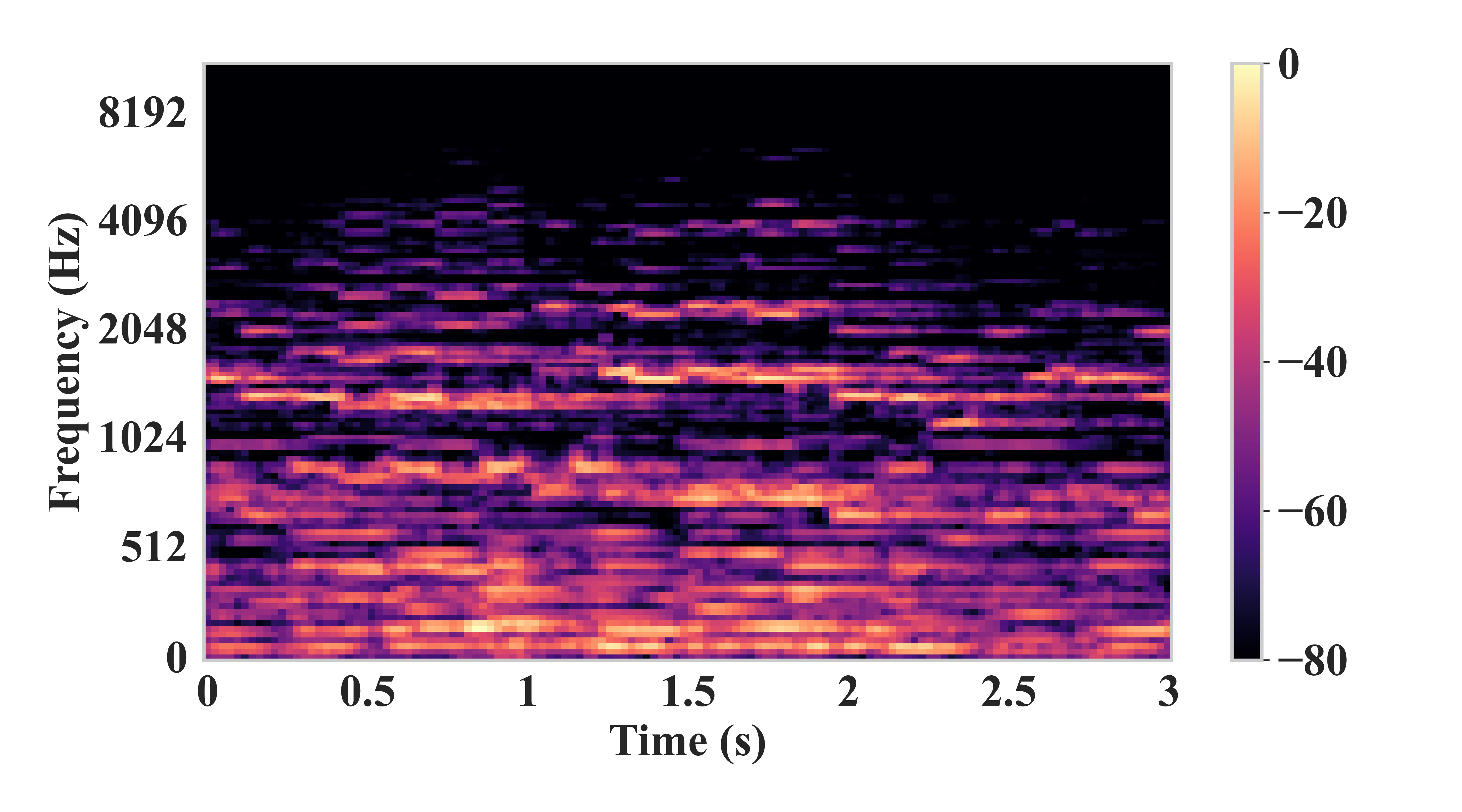}
  \caption{Mel spectrogram of an audio sample.}
  \label{fig:mel}
\end{figure}
Additionally, numerous subscription and recommendation systems require the inclusion of music genre preferences to provide customers with more specific and tailored content. Music is a highly diverse art form that encompasses various elements, including rhythm, melody, and harmony. Music media systems use textual labels to categorize or retrieve music, since comprehending music in its exact form requires substantial previous knowledge. Hence, the classification of music genres is an essential component of music information retrieval ~\cite{singh2022robustness}.

\begin{figure*}[ht!]
  \centering
  \includegraphics[width= .88\textwidth]{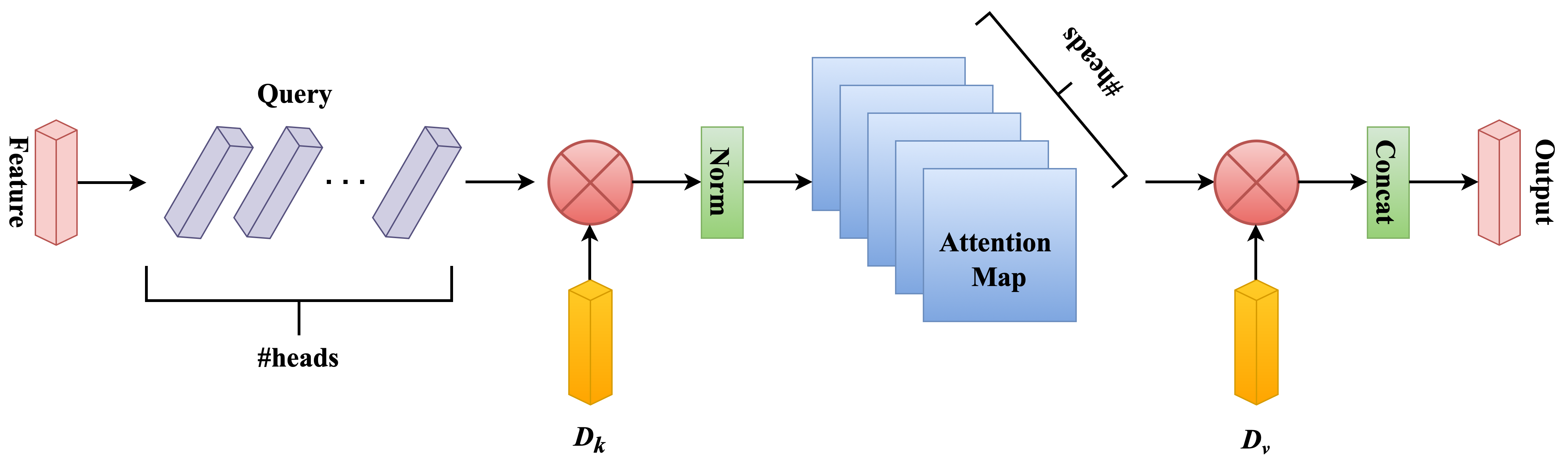}
  \caption{ Multi-head external-attention (MEA).}
  \label{fig:onecol}
\end{figure*}

Over the past decade, traditional machine learning classifiers have been employed for this purpose. However, these classifiers possess a shallow structure, limiting their capacity to effectively learn and interpret music data. Manual extraction of musical elements has shown limited applicability and lacks robustness. To address the intricate aspects of music, innovative deep-learning approaches have been adopted in recent years. These advanced techniques offer improved accuracy and efficiency in music genre classification and retrieval~\cite{prabhakar2023holistic, sterling2018isnn}. The most recent advancements in audio classification performance are generated using distinct resource-intensive methodologies~\cite{zhang2019learning}. 

Over the years, a large number of classification methods have been created to categorize audio-based data~\cite{hershey2017cnn}. 
A study conducted video classification and audio event identification using a convolutional neural network (CNN) on an extensive dataset. By utilizing subsets of varying sizes and initializing convolutional layers with a gammatone filter bank, which is modelled after the human ear, the study examined the impact of these methods on the system's accuracy. This approach allowed for a thorough investigation of model training and performance~\cite{mishachandar2021diverse}. Researchers have employed Long Short-Term Memory (LSTM) networks to achieve promising results for the MGC task~\cite{dai2016long}. Another study introduced a hybrid architecture combining CNNs and RNNs to capture the spatiotemporal characteristics of music~\cite{choi2017convolutional}. Additionally, a study proposed an attention mechanism-based Bi-LSTM architecture to exploit the varying importance of temporal frames, further enhancing the accuracy and performance of music classification tasks~\cite{yu2020deep}.
A research study compared a CNN and a TDSN for environmental sound classification using spectrogram images. The results revealed that the CNN achieved superior performance on the ESC-10 dataset, with an accuracy of 77\%, compared to 56\% for the TDSN ~\cite{khamparia2019sound}. Additionally, a lightweight CNN utilizing MFCC for environmental sound classification on the UrbanSound8k dataset achieved a notable accuracy of 95.59\%, showcasing competitive performance against more complex models~\cite{al2021rethinking}. A further investigation concentrated on classifying music across three distinct datasets (ISMIR 2004, LMD, and ethnic African music). The findings revealed that CNN outperformed previous classifiers by achieving an impressive accuracy rate of 92\% ~\cite{costa2017evaluation}.

The recently introduced ViT offers a substantial advancement in using transformer attention models for computer vision applications, exhibiting notable enhancements in performance compared to current models ~\cite{dosovitskiy2020image}. Recent advancements include the Causal Audio Transformer (CAT), which employs MRMF extraction and an acoustic attention block for enhanced audio modelling. The proposed causal module mitigates overfitting, facilitates knowledge transfer, and improves interpretability. It demonstrated superior performance on datasets such as ESC50, AudioSet, and UrbanSound8K, and are adaptable to various other transformer-based models~\cite{liu2023cat}. Another study applied a transformer model to music genre classification, converting audio to log-amplitude Mel spectrograms, and achieved promising results on the GTZAN dataset, indicating the potential of transformers in this domain~\cite{zhuang2020music}.

In this study, EAViT with MEA is proposed to classify music genres in the GTZAN dataset. The proposed approach includes the introduction of an innovative EA mechanism to enhance the classification accuracy of ViT-based models when applied to audio data. EAViT integrates the strengths of traditional vision transformers with specialized attention mechanisms tailored for audio signal processing, achieving superior performance over conventional models. Extensive experiments were conducted on the GTZAN dataset to benchmark our model against state-of-the-art methods, demonstrating significant improvements in classification accuracy. Our contributions in this paper are summarized as follows: 

\begin{itemize}
    \item EAViT is proposed for music genre classification. EAViT utilizes MEA in its process.
    \item A 30-second raw audio clip is segmented into 3-second portions. Each segment is transformed into the frequency domain via Short-Time Fourier Transform (STFT), converted to decibel units to produce a spectrogram, and further processed with a mel filter bank to create a mel spectrogram, which is then saved as an image. 
    \item The proposed models are compared with state-of-the-art methods from existing literature.

\end{itemize}
 
The rest of the paper is structured as follows: Section II outlines the methodology, Section III presents the results, and Section IV contains the conclusion.

\section{Methodology}

\subsection{Data}
\begin{figure*}[ht!]
  \centering
  \includegraphics[width= .97\textwidth]{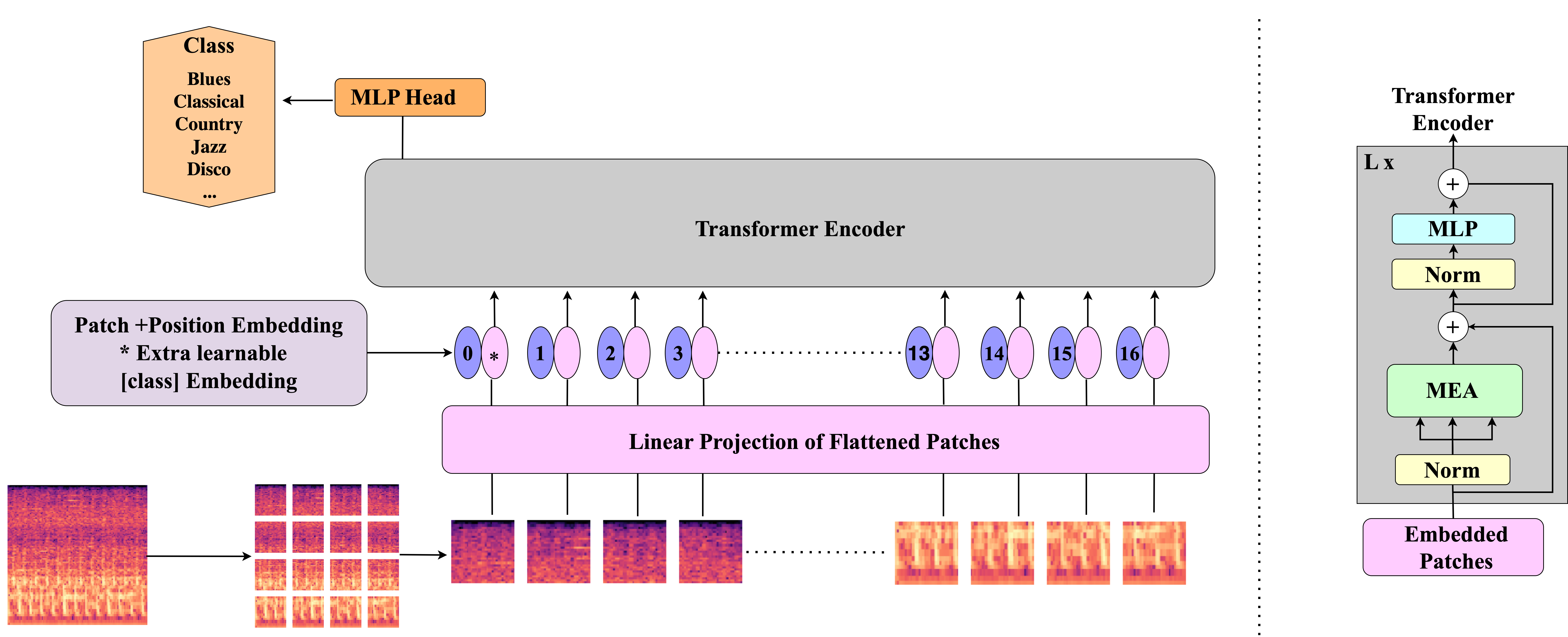}
  \caption{Model Overview: EAViT.}
  \label{fig:EAVIT}
\end{figure*}

GTZAN dataset was used which comprises of 1,000 audio excerpts spanning 10 different genres: Blues, Classical, Country, Disco, Hip-hop, Jazz, Metal, Pop, Reggae, and Rock. Each genre includes 100 excerpts, each lasting approximately 30 seconds and stored as 22,050 Hz, 16-bit, mono audio files~\cite{tzanetakis2002musical}.  

We segmented each 30-second audio excerpt from the GTZAN dataset into 3-second clips, significantly expanding the dataset and mitigating the risk of overfitting. This segmentation allows for a more granular analysis of audio features, enhancing the dataset's robustness. To illustrate the transformation process, we included both the raw audio signals and the corresponding mel spectrogram Fig.~\ref{fig:raw} and Fig.~\ref{fig:mel}.

Mel Spectrogram represents sound in the time-frequency domain, using the mel scale on the y-axis and the decibel scale to indicate colour intensity. This spectrogram is created by applying a filter bank to the frequency-domain representation of time-windowed audio signals, providing a perception-aligned view of the sound. The conversion process involves transforming each 3-second audio segment into the frequency domain using the STFT. The resulting frequency data is then converted into decibel units to form a standard spectrogram. By applying a mel filter bank, we transform this spectrogram into a mel spectrogram, which is then saved as an image file.

\subsection{Multi-head external-attention (MEA)}

In this study, we incorporated the MEA mechanism into the transformer encoder of the ViT. To improve the network's capabilities, the EA mechanism uses two learnable memory units, \( M_k \in \mathbb{R}^{d \times S} \) and \( M_v \in \mathbb{R}^{d \times S} \), which serve as key and value components. Importantly, \( M_k \) and \( M_v \) function as memories for all samples in the training set and are independent of the input. Specifically, the flow process of the EA module is defined by Eq.(\ref{eq:1}) and (\ref{eq:2}), as follows:

\begin{equation}
A = (\alpha)_{n,m} = \text{Norm}(F_{\text{in}} M_k^T),
\label{eq:1}
\end{equation}

\begin{equation}
F_{\text{out}} = A M_v,
\label{eq:2}
\end{equation}

where \( F_{\text{in}} \) represents the input feature, \( F_{\text{out}} \) represents the output feature, and \(\text{Norm}(\cdot)\) denotes the double normalization operator. MEA can be written as:

\begin{align}
h_i &= \text{ExternalAttention}(F_i, M_k, M_v), \\
F_{\text{out}} &= \text{MultiHead}(F, M_k, M_v) \\
&= \text{Concat}(h_1, \ldots, h_H) W_o,
\end{align}

where \( h_i \) represents the \( i \)-th head, \( H \) denotes the total number of heads, and \( W_o \) is a linear transformation matrix ensuring consistent input and output dimensions. \( M_k \in \mathbb{R}^{d \times S} \) and \( M_v \in \mathbb{R}^{d \times S} \) serve as the shared memory units across different heads~\cite{guo2022beyond}. Here Fig.~\ref{fig:onecol} shows the structure MEA.

\section{Proposed Method}

\subsection{ External Attention Vision Transformer (EAViT)}

 The conventional ViT utilizes multi-head self-attention, which enhances the feature at each position by computing a weighted sum of features based on pair-wise affinities across all positions. This process effectively captures long-range dependencies within a single sample. However, self-attention suffers from quadratic complexity and overlooks potential correlations between different samples. In contrast, EA leverages two small, learnable, shared memories, which can be efficiently implemented using two cascaded linear layers and two normalization layers. In this study, we have utilized the EAViT model, which incorporates MEA within the transformer encoder. 
 
 The ViT is specifically designed for image classification tasks by directly applying the transformer architecture to sequences of image patches~\cite{dosovitskiy2020image}. It closely adheres to the original transformer design. In order to process 2D images, the input tensor $x\in R^{H\times W\times C}$ is transformed into a set of flattened 2D patches denoted as $x_{P}\in R^{N\times (P^{2}\cdot C)}$, where $C$ represents the number of channels. The main image has a resolution of $(H,W)$, while each image patch has a $(P,P)$ resolution. Consequently, the actual sequence length for the transformer is given by $N=HW / P^2$. The output of a learnable linear projection (as expressed in Eq.(\ref{eq:one})) called patch embeddings, associates each vectorized patch with the model dimension $D$, as the transformer maintains consistent widths across all its layers.

\begin{equation}\label{eq:one}
\mathrm{z}_{0} =\left[\mathrm{x}_{\text {class }} ; \mathrm{x}_{p}^{1} \mathbf{E}~ ; \mathrm{x}_{p}^{2} \mathbf{E}~; \ldots~; \mathrm{x}_{p}^{N} \mathbf{E}\right]+\mathbf{E}_{p o s},  
\end{equation}
where $\mathbf{E} \in \mathbb{R}^{\left(P^{2} \cdot C\right) \times D} $ and $ \mathbf{E}_{\text {pos}} \in \mathbb{R}^{(N+1) \times D}.   $

Researchers introduce a trainable embedding into the sequence of embedded patches, akin to BERT's $[class]$ token, represented as $(z_0^0 = x_{\text{class}})$. This embedding's state at the output of the Transformer encoder, denoted as $(z_L^0)$, is used as the image representation $y$ (refer to Eq.(\ref{eq:four})). During both the pre-training and fine-tuning stages, a classification head is connected to $z_L^0$. In the pre-training phase, a Multi-Layer Perceptron (MLP) with a single hidden layer guides the classification head. In contrast, a single linear layer is utilized during the fine-tuning phase ~\cite{zim2022vision}. Here Fig.~\ref{fig:EAVIT} shows the overview of the model.

\begin{equation}\label{eq:two}
\mathrm{z}_{\ell}^{\prime} =\operatorname{MEA}\left(\operatorname{LN}\left(\mathrm{z}_{\ell-1}\right)\right)+\mathrm{z}_{\ell-1}, \quad \ell=1 \ldots L    
\end{equation}
\begin{equation}\label{eq:three}
\mathrm{z}_{\ell} =\operatorname{MLP}\left(\operatorname{LN}\left(\mathrm{z}_{\ell}^{\prime}\right)\right)+\mathrm{z}_{\ell}^{\prime}, \quad \ell=1 \ldots L    
\end{equation}
\begin{equation}\label{eq:four}
\mathrm{y} =\operatorname{LN}\left(\mathrm{z}_{L}^{0}\right)    
\end{equation}

To retain positional information, position embeddings are added to the patch embeddings. As more complex 2D-aware position embeddings did not significantly enhance performance, the researcher opted for traditional learnable 1D position embeddings. The generated sequence of embedding vectors is then fed into the encoder. The Transformer encoder comprises layers of MEA and MLP blocks, as described in Eq.(\ref{eq:two}) and (\ref{eq:three}). Each block is preceded by Layernorm (LN) and followed by residual connections. 

In our proposed transformer-based methodology for the classification of 2D Mel spectrogram images, critical hyperparameters are meticulously selected to enhance model efficacy. The input images, resized to $256\times 256$ pixels, are segmented into 16 patches. The learning rate is established at 0.001, accompanied by a weight decay of 0.0001 to mitigate overfitting, and the batch size is set at 256 to facilitate efficient training iterations. The transformer architecture incorporates 16 layers, each with a projection dimension of 32 and 8 attention heads. The MLP classification head comprises two hidden layers with [2048, 1024] units. These hyperparameters collectively define the model's capacity, attentiveness, and overall classification performance.
\footnotetext[7] {The model has been reimplemented.}

\begin{table}[htbp]
    \caption{Proposed models comparison with SOTA methods from literature.}
    \begin{center}
        \begin{tabular}{|c|c|}
            \hline
            \textbf{Model} & \textbf{Accuracy} \\
            \hline
            ViT\footnotemark [7] \ & 91.79\% \\
            \hline
             RNNCA~\cite{gan2021music} & 93.1\% \\
            \hline
            1D-CNN with BiRNN and attention mechanism~\cite{zhang2021music} & 91.99\% \\
            \hline
            SA-SLnO with optimization~\cite{kumaraswamy2021deep} & 85.63\% \\
            \hline
            \textbf{EAViT (our)} & \textbf{93.99\%} \\
            \hline
        \end{tabular}
        \label{tab:accuracy}
    \end{center}
\end{table}
\begin{table}[htbp]
\caption{Evaluation of class-specific classification results of the proposed EAViT model}
\begin{center}
\begin{tabular}{|c|c|c|c|}
\hline
\textbf{Class} & \textbf{Precision} & \textbf{Recall} & \textbf{F1-Score} \\
\hline
Blues & 0.94 & 0.96 & 0.95 \\
\hline
Classical & 0.99 & 0.97 & 0.98 \\
\hline
Country & 0.89 & 0.93 & 0.91 \\
\hline
Disco & 0.95 & 0.97 & 0.96 \\
\hline
Hiphop & 0.93 & 0.90 & 0.91 \\
\hline
Jazz & 0.92 & 0.94 & 0.93 \\
\hline
Metal & 0.96 & 0.93 & 0.95 \\
\hline
Pop & 0.99 & 0.92 & 0.96 \\
\hline
Reggae & 0.93 & 0.92 & 0.93 \\
\hline
Rock & 0.90 & 0.94 & 0.92 \\
\hline
\end{tabular}
\label{tab:results}
\end{center}
\end{table}

\begin{figure}[t]
  \centering
  \includegraphics[width= 0.38\textwidth]{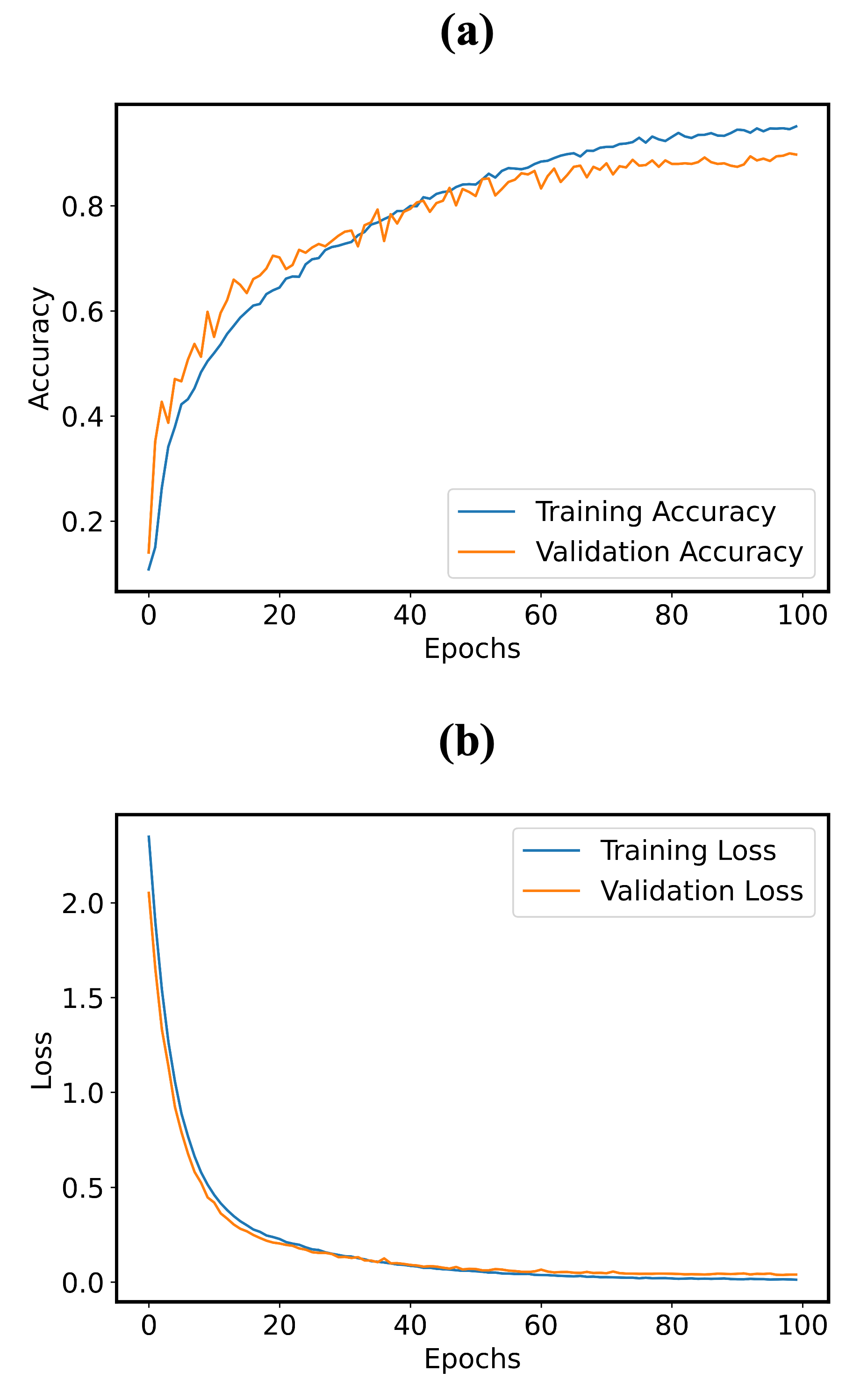}
  \caption{proposed models: (a) Training and validation accuracy over epochs. (b) Training and validation loss over epochs. }
  \label{fig:acc_loss}
\end{figure}

\begin{figure}[t]
  \centering
  \includegraphics[width= 0.48\textwidth]{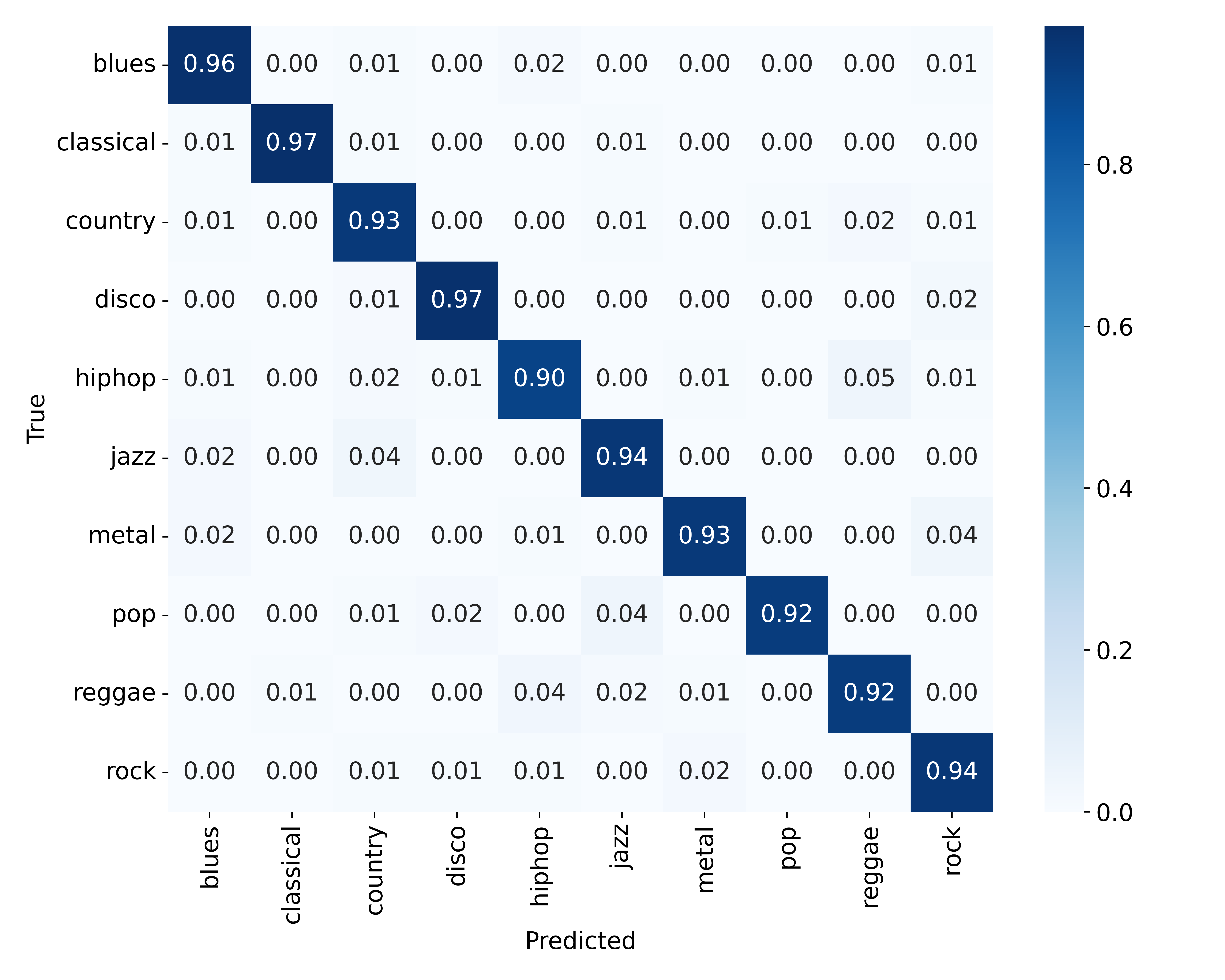}
  \caption{Confusion matrix of the model.}
  \label{fig:con_matrix}
\end{figure}

\section{Experiments And Results}

\subsection{Evaluation Matrices}

To evaluate the performance of the models, we employed key performance metrics, including accuracy, precision, recall, and the F1-score. Accuracy represents the ratio of correct predictions to the total number of predictions. Precision is calculated as the ratio of true positive predictions to the sum of true positive and false-positive predictions across all classes. Recall is the ratio of true positive predictions to the sum of true positive and false-negative predictions across all classes. The F1-Score is the harmonic mean of precision and recall, balancing the consideration of both false positive and false negative predictions. The mathematical expressions for these metrics are as follows:

\begin{equation}
\text{Accuracy} = \frac{TP + TN}{TP + FP + TN + FN} \times 100\%
\end{equation}

\begin{equation}
\text{Precision} = \frac{TP}{TP + FP}
\end{equation}

\begin{equation}
\text{Recall} = \frac{TP}{TP + FN}
\end{equation}

\begin{equation}
\text{F1-score} = 2 \times \left(\frac{\text{Recall} \times \text{Precision}}{\text{Recall} + \text{Precision}}\right)
\end{equation}

Where TP denotes True Positives, TN denotes True Negatives, FN denotes False Negatives, and FP denotes False Positives.

\subsection{Environment}

The experiment was conducted using Python 3.11. The hardware employed for both training and testing included an AMD Ryzen 9 5900X processor, 64GB of RAM, and an Nvidia GeForce RTX 3090 GPU with 24GB of VRAM.

\subsection{Analysis}

In this study, the EAViT model was employed to classify ten different types of music genres. The GTZAN dataset was used in this study. EAViT achieved an impressive overall accuracy of 93.99\%. The results underscore the exceptional performance of the EAViT model in music genre classification. Here, Fig.~\ref{fig:acc_loss} (a) illustrates the training and validation accuracy, and Fig.~\ref{fig:acc_loss} (b) presents the training and validation loss over a span of 100 epochs respectively. The graphs indicate that the model demonstrates optimal fitting to the dataset. The confusion matrix depicted in Fig.~\ref{fig:con_matrix} provides a comprehensive overview of the model's performance, illustrating the models classification accuracy. 

To demonstrate the robustness and efficacy of the proposed model in this study, we conducted a comparative analysis with state-of-the-art (SOTA) models from the literature, including ViT~\cite{dosovitskiy2020image}, Recurrent Neural Networks with Channel Attention Mechanism (RNNCA)~\cite{gan2021music}, 1D-CNN with BiRNN and attention mechanism~\cite{zhang2021music}, and SA-SLnO with optimization~\cite{kumaraswamy2021deep}. Table~\ref{tab:accuracy} unequivocally shows that the results of our proposed method surpass the performance of these SOTA models. EAViT surpasses the ViT by 2.37\%, RNNCA by 0.95\%, and the 1D-CNN with BiRNN and attention mechanism by 2.15\%. Furthermore, in comparison to SA-SLnO with optimization model, EAViT achieves a remarkable 9.31\% increase in classification accuracy. Here, Table~\ref{tab:results} presents the evaluation metrics, including Precision, Recall, and F1-score, for each class of the proposed EAViT model. These findings underscore the robustness of the proposed model as it demonstrates consistent performance across various classes.

\section{Conclusion}

In this study, we introduced the External Attention Vision Transformer (EAViT) model, designed to enhance audio classification accuracy by integrating MEA mechanisms within the ViT framework. The EAViT model effectively captures long-range dependencies and correlations between different audio samples, addressing the limitations of traditional self-attention mechanisms. Through extensive experimentation on the GTZAN dataset, comprising 1,000 audio excerpts across ten genres, EAViT achieved a notable accuracy of 93.99\%, outperforming state-of-the-art models such. Key metrics, including precision, recall, and F1-score, further demonstrated the model's robustness and reliability. The segmentation of 30-second audio clips into 3-second excerpts enhanced dataset robustness and allowed for detailed feature analysis. These results underscore the potential of the EAViT model to significantly advance audio classification, improving user experiences in various audio-related applications, including music streaming and environmental sound recognition. Future work could explore EAViT's application to other audio classification tasks and datasets, highlighting its versatility and efficacy.

\bibliographystyle{IEEEtran}
\bibliography{text} 

\end{document}